\documentclass{iau}

\usepackage{amsmath}
\usepackage{graphicx}
\usepackage{multirow}

\newcommand{\mnras}{MNRAS}
\newcommand{\apj}{ApJ}

\newcommand{\apjs}{ApJS}
\newcommand{\aj}{AJ}
\newcommand{\nat}{Nature}

\newcommand{\pasp}{PASP}

\begin{document}

\lefttitle{P. Odert et al.}
\righttitle{Spectroscopy of flares on AU Mic}

\jnlPage{1}{7}
\jnlDoiYr{2024}
\doival{10.1017/xxxxx}
\volno{388}
\pubYr{2024}
\journaltitle{Solar and Stellar Coronal Mass Ejections}

\aopheadtitle{Proceedings of the IAU Symposium}
\editors{N. Gopalswamy,  O. Malandraki, A. Vidotto \&  W. Manchester, eds.}

\title{Spectroscopy of flares on AU Mic}

\author{P. Odert$^1$, M. Leitzinger$^1$, P. Kabáth$^2$, R. Greimel$^3$, J. Lipták$^{2,4}$, R. Karjalainen$^2$,\\ P. Heinzel$^{2,5}$, J. Wollmann$^2$, E.W. Guenther$^6$}
\affiliation{$^1$Institute of Physics/AGP, University of Graz, Graz, Austria}
\affiliation{$^2$Astronomical Institute of the Czech Academy of Sciences, Ondřejov, Czech Republic}
\affiliation{$^3$RG Science, Graz, Austria}
\affiliation{$^4$Astronomical Institute of Charles University, Prague, Czech Republic}
\affiliation{$^5$Center of Scientific Excellence - Solar and Stellar Activity, University of Wrocław, Wrocław, Poland}
\affiliation{$^6$Thüringer Landessternwarte Tautenburg, Tautenburg, Germany}

\begin{abstract}
Photometric space missions like Kepler and TESS have revealed unprecedented statistics of stellar flares, including the frequent detection of highly energetic superflares. Spectroscopic characterization of flares and especially superflares, however, is still not as commonly available. In contrast to broad-band photometry, spectroscopic data can give more insights into the physical parameters of flares, as well as allowing to identify possibly accompanying CMEs. In this context, we monitored the young planet-hosting flare star AU~Mic for about 60 partial nights ($>$160 hours) between 2022 and 2023 spectroscopically with the ESO 1.52m telescope hosted by the PLATOSpec consortium. Here, we report on the search for signatures of flares and CMEs in these data, including the partial observation of a possible extreme superflare.
\end{abstract}


\maketitle

\section{Introduction}
AU~Mic is a young early M dwarf hosting several planets \citep{Plavchan20, Wittrock23, Donati23} and a debris disk \citep{Kalas04}. Due to its young age \citep[23\,Myr;][]{Mamajek14}, the star is magnetically very active, possessing strong magnetic fields \citep{Donati23} and stellar spots \citep{Ikuta23}. It exhibits frequent, strong flares and even superflares \citep{Gilbert22, Ikuta23, Tristan23}. Superflares are flares with energies $>10^{33}$\,erg \citep{Maehara12}, mainly detected and studied using broad-band photometry. Spectroscopic investigations of flares and superflares may provide more information on their physical properties and generation mechanisms than what can be extracted from broad-band photometry, but are far more time-consuming to obtain. Therefore, due to its high flare rate, AU~Mic provides an ideal target to characterize its flares and superflares spectroscopically, which may also provide insights into the space weather environment of its planets.

\section{Observations}
Spectroscopic monitoring of AU~Mic was performed with the ESO1.52m telescope (E152) located in La~Silla, Chile, which is operated by the PLATOSpec consortium\footnote{The PLATOSpec consortium is led by the Astronomical Institute of the Czech Academy of Sciences and consisting of Thüringer Landessternwarte Tautenburg and Universidad Catolica de Chile as major partners and of Universidad Adolfo Ibanez and Masaryk University as minor partners and University of Graz as collaborating partner (\url{https://stel.asu.cas.cz/plato/}).}. The telescope is currently equipped with the Echelle spectrograph PUCHEROS+, an upgraded version of PUCHEROS \citep{Vanzi12}. Its resolving power is R$\approx$18000 and it provides a wavelength coverage of about 400 to 700\,nm. In addition, the E152 is equipped with two 15\,cm finder telescopes, on which two cameras are installed. These can be used for simultaneous photometry and/or low-resolution slitless spectroscopy.

We observed AU~Mic for 56 partial nights between 2022-10-31 and 2023-09-21 with PUCHEROS+. For 38 of these nights we obtained simultaneous photometry in the g'-band, and for two nights simultaneous low-resolution spectroscopy. Exposure times vary between 5, 10, and 15\,min for the Echelle spectroscopy, and are typically a few seconds for the photometry. The PUCHEROS+ data are reduced with the CERES+ pipeline, which is an updated version of CERES \citep{Brahm17} and generates wavelength-calibrated, optimally extracted order-by-order spectra. More details on the instruments and the observing campaign can be found in \citet{Odert25}.

\section{Flare detection}
\subsection{Spectroscopy}

\begin{figure}[t]
  \centerline{\vbox to 6pc{\hbox to 10pc{}}}
  \includegraphics[height=5cm]{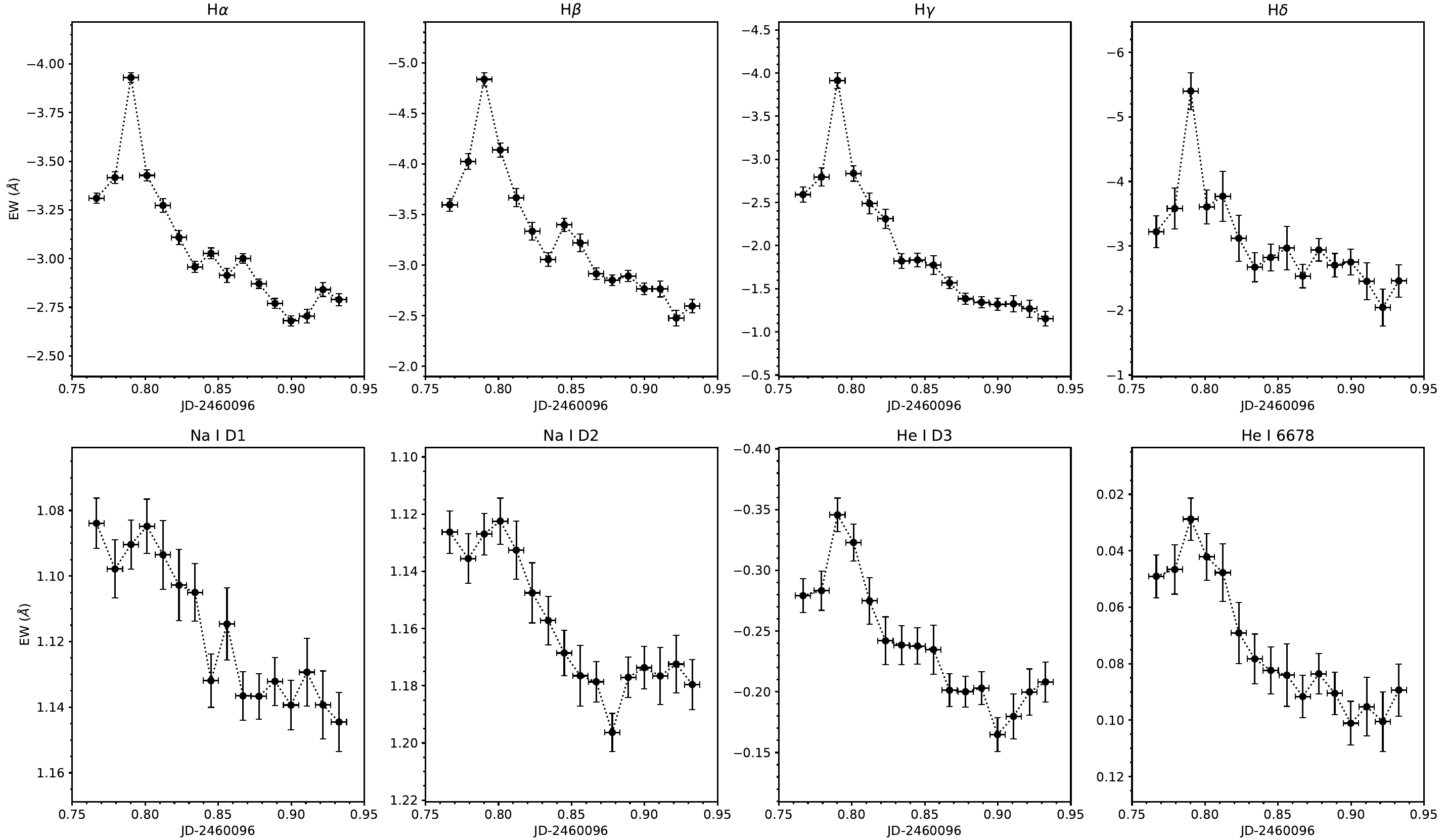}
  \hfill
  \includegraphics[height=5cm]{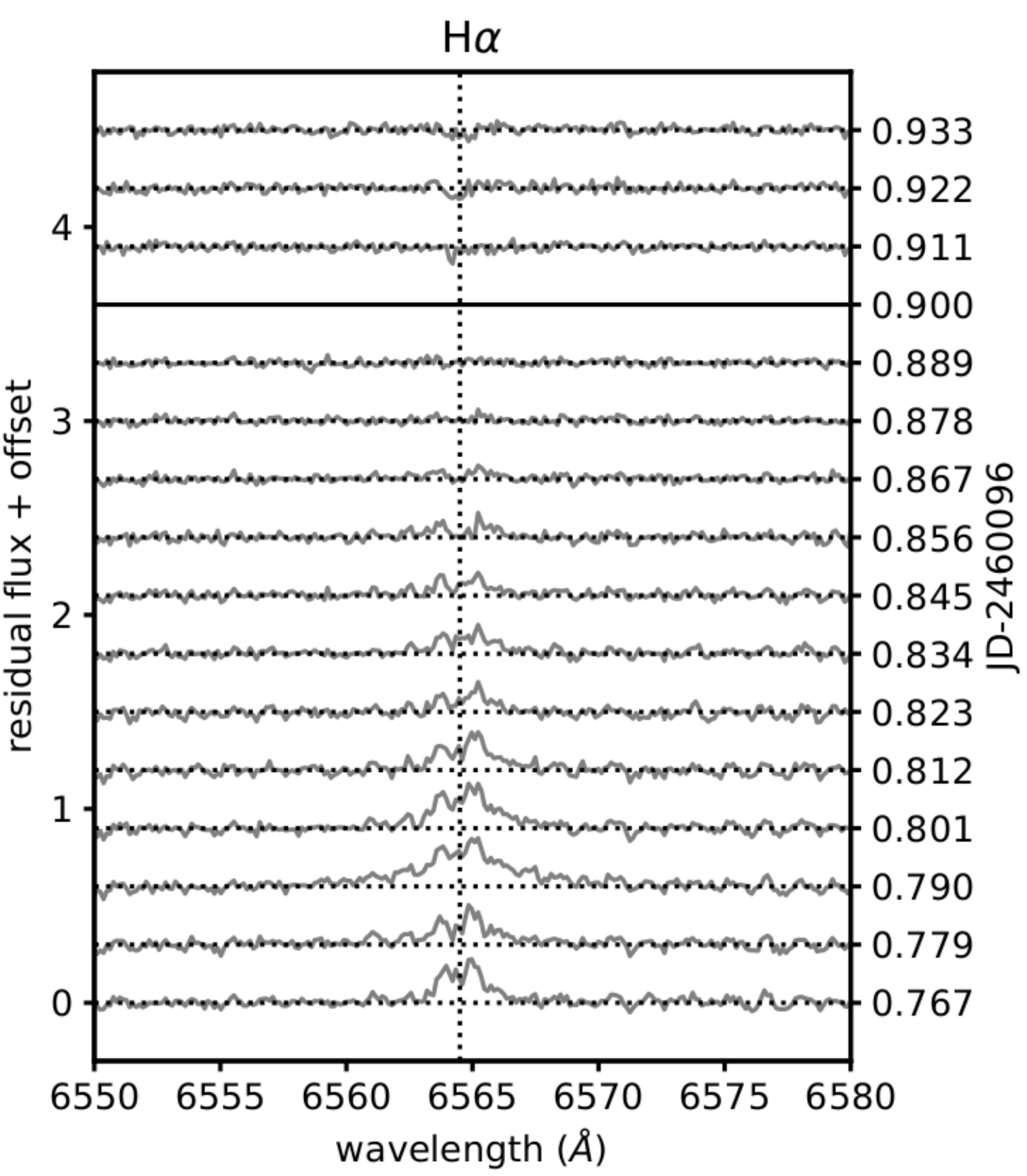}
  \caption{Left panels: Flare on 2023-06-01 in different chromospheric lines. Right panel: corresponding time series of residual spectra around the H$\alpha$ line.}
  \label{fig:specflare}
\end{figure}

To detect flares in the spectroscopic data, we generate light curves of the equivalent widths (EWs) of several chromospheric lines (H$\alpha$, H$\beta$, H$\gamma$, H$\delta$, Na\,I\,D1\&D2, He\,I\,D3, He\,I\,6678). We identify 24 flares in the light curves of these lines, an example from 2023-06-01 is shown in Fig.\,\ref{fig:specflare}. One of the detected flares is a partial observation of an extreme event, which is described in more detail in Section\,\ref{sec:superflare}. Excluding this event, the other 23 flares have H$\alpha$ peak luminosities ranging from $\sim10^{27}$ to $\sim10^{28}$\,erg\,s$^{-1}$ and H$\alpha$ energies between $\sim3\times10^{30}$ and $\sim5\times10^{31}$\,erg.

A currently ongoing task is the search for filament/prominence eruptions as indicator for stellar CME activity. Therefore, we aim to identify spectral line asymmetries to find transient blue- or red-shifted line components. An example is shown in the right panel of Fig.\,\ref{fig:specflare}, which shows the residual spectra around H$\alpha$ for the flare event on 2023-06-01 displayed in the left panels. It reveals a small red asymmetry (excess emission) around the peak of the flare. 

\subsection{Photometry}
\begin{figure}[t]
  \centerline{\vbox to 6pc{\hbox to 10pc{}}}
  \includegraphics[height=4cm]{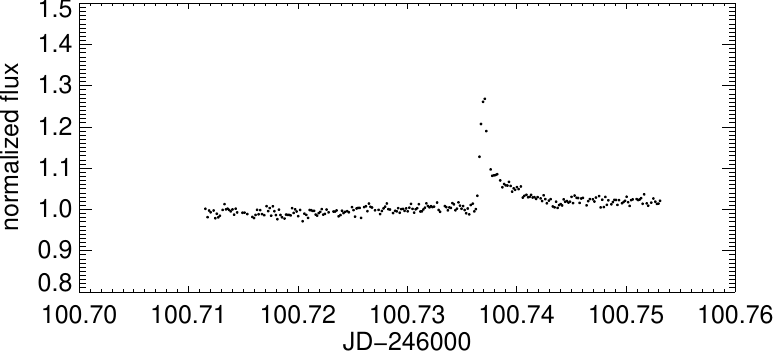}
  \hfill
  \includegraphics[height=4.5cm]{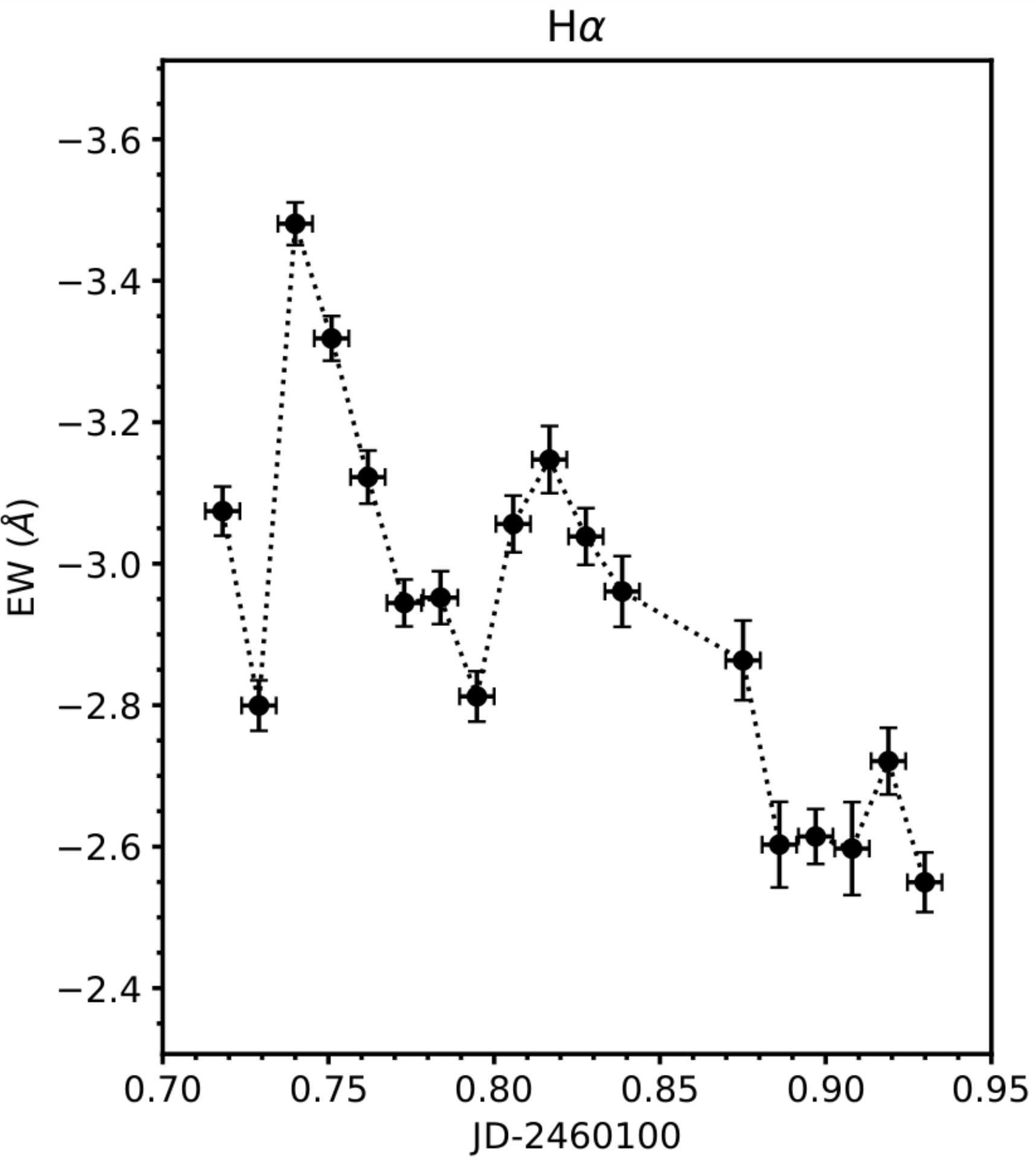}
  \caption{Left panel: g'-band flare on 2023-06-05. Right panel: associated flare in H$\alpha$.}
  \label{fig:gbandflare}
\end{figure}

From the accompanying photometric data we detect one g'-band flare (Fig.\,\ref{fig:gbandflare}), associated with the most energetic H$\alpha$ event for which we had simultaneous photometric observations of sufficient quality. The flare has an amplitude of $\sim$27\% and lasts for about 11\,min. Its estimated energy is $\sim10^{32}$\,erg in the g'-band, and we extrapolate a total white-light energy of $\sim10^{33}$\,erg following \citet{Shibayama13}, which makes it a superflare. The associated flare detected in the chromospheric lines lasts much longer, about 290\,min. Its energy in H$\alpha$ amounts to $\sim4\times10^{31}$\,erg, and in total to $\sim10^{32}$\,erg in all studied chromospheric spectral lines.

\section{Extreme event}
\label{sec:superflare}
\begin{figure}[h]
  \centerline{\vbox to 6pc{\hbox to 10pc{}}}
  \includegraphics[scale=.25]{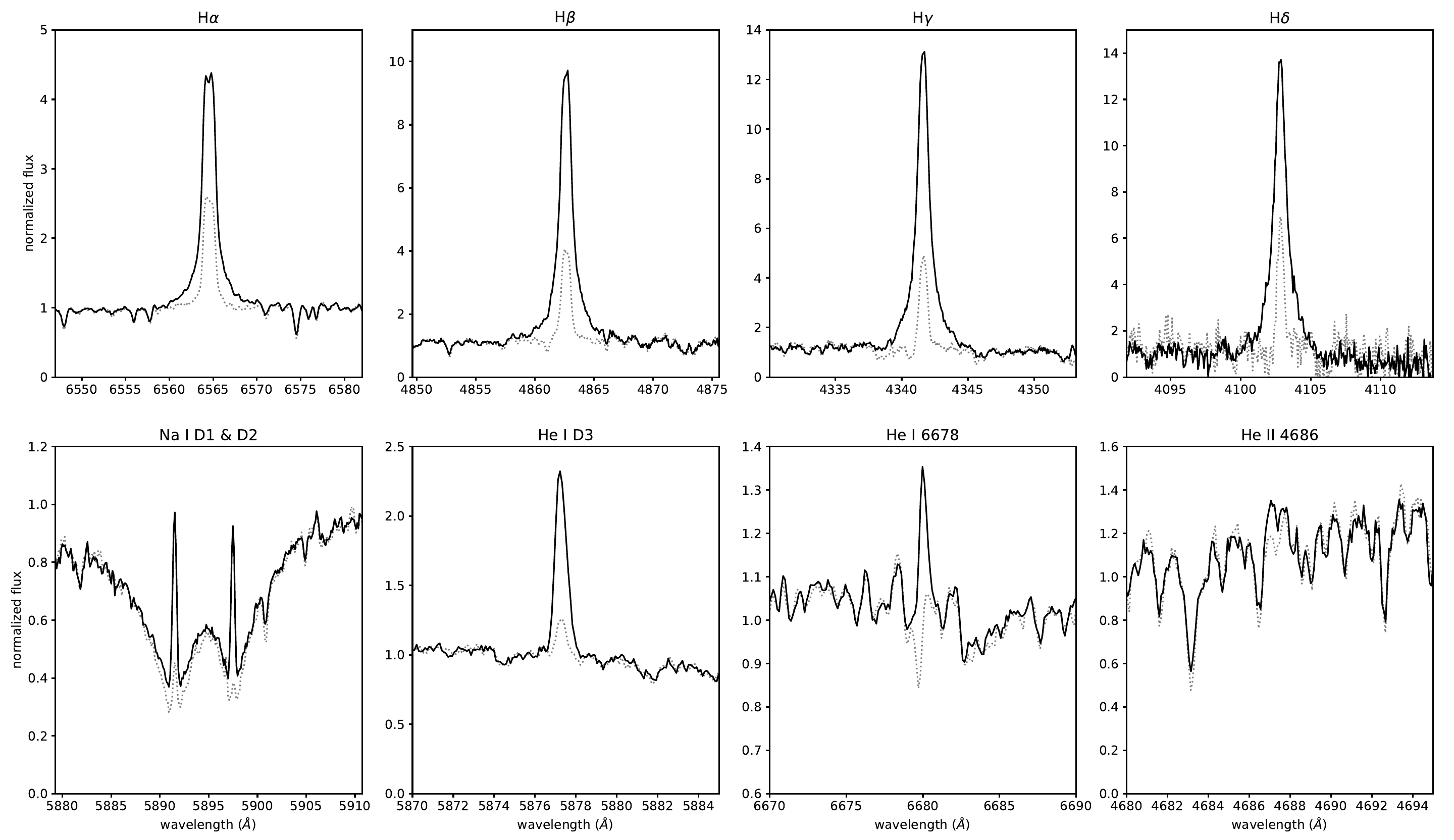}
  \caption{Comparison of spectral line profiles of the extreme event on 2023-09-16 (black solid line) compared with a low-activity spectrum from the following night 2023-09-17 (gray dotted line).}
  \label{fig:superflare}
\end{figure}

On 2023-09-16 we observed an extreme event with line fluxes increasing by factors of a few more than in any other night (see Fig.\,\ref{fig:superflare}). Unfortunately, the observations covered only about 30\,min, during which all line fluxes remained at a roughly similar level of enhancement. Thus, only a small part of a likely much longer event was observed. Upon close inspection of the EW light curves of the following night (2023-09-17), a decreasing trend is seen in the data. Possibly, this could resemble the decaying tail of the extreme event of 2023-09-16. If true, this would indicate a duration of this giant flare of more than 24\,hours. We fit the combined EWs from the two nights for each spectral line with a simple flare light curve model and integrate over the fits to obtain an estimate of the energies. This results in superflare energies ($\gtrsim10^{33}$\,erg) already in H$\alpha$ alone. To independently check this assumption, we utilize the relationship between flare peak luminosity and energy derived from the sample of AU~Mic's other 23 flares we detected in our study \citep{Odert25}. If we take the flare luminosity measured on 2023-09-16 as the peak luminosity, we obtain flare energies comparable to within a factor of two with our estimates from the light curve fits. This agreement substantiates our hypothesis of a rare, highly-energetic long duration superflare.

In addition to the prominent chromospheric lines presented here, we find a wealth of other spectral lines which were significantly enhanced or appeared during this extreme event. One example is the He\,II line at 4686\,\AA\ (see Fig.\,\ref{fig:superflare}, lower right panel). A detailed analysis of flare-affected lines of this event will be the subject of a future study.

\section{Conclusions and outlook}
We present results from a spectroscopic monitoring campaign of the active young planet host star AU~Mic. From the spectral line data, we identified 24 flares, including the partial observation of a rare extreme event. A detailed analysis of flare-affected spectral lines and line asymmetries, also comparatively between weaker flares and superflares, is currently ongoing.

\vspace{\baselineskip}

\noindent\textbf{Acknowledgments}\\
The authors acknowledge the Austrian Science Fund (FWF): 10.55776/I5711 and the Czech Science Foundation (GACR): 22-30516K. PH was supported by the program 'Excellence Initiative - Research University' for years 2020-2026 at University of Wroc{\l}aw, project No. BPIDUB.4610.96.2021.KG.

\end{document}